\documentclass[pdflatex,sn-mathphys-num]{sn-jnl}% Math and Physical Sciences Numbered Reference Style
\usepackage{graphicx}%
\usepackage{multirow}%
\usepackage{amsmath,amssymb,amsfonts}%
\usepackage{amsthm}%
\usepackage{mathrsfs}%
\usepackage[title]{appendix}%
\usepackage{xcolor}%
\usepackage{textcomp}%
\usepackage{manyfoot}%
\usepackage{booktabs}%
\usepackage{algorithm}%
\usepackage{algorithmicx}%
\usepackage{algpseudocode}%
\usepackage{listings}%
\usepackage{bm}
\usepackage{siunitx}
\theoremstyle{thmstyleone}%
\theoremstyle{thmstyletwo}%

\theoremstyle{thmstylethree}%

\begin{document}

\title[Article Title]{Coarse Graining Reveals a Fluctuation-theorem-like Asymmetry in Financial Markets}


\author[1]{\fnm{Jian} \sur{Gao}}\email{gao.jian@bupt.edu.cn}
\equalcont{These authors contributed equally to this work.}

\author[1]{\fnm{Lufeng} \sur{Zhang}}\email{lfzhang@bupt.edu.cn}
\equalcont{These authors contributed equally to this work.}

\author*[1]{\fnm{Ping} \sur{Fang}}\email{pingfang@bupt.edu.cn}

\author[2,3]{\fnm{Pu} \sur{Ke}}\email{kepu@superbloch.com}

\author[3]{\fnm{Jin} \sur{Wu}}\email{wujin\_astro@pku.edu.cn}

\author[4]{\fnm{Yue} \sur{Liu}}\email{yue.liu@yukawa.kyoto-u.ac.jp}

\author[5,6,7]{\fnm{Haijun} \sur{Zhou}}\email{zhouhj@itp.ac.cn}

\affil[1]{\orgdiv{School of Physical Sciences and Technology}, \orgname{Beijing University of Posts and Telecommunications}, \orgaddress{\postcode{100876}, \state{Beijing}, \country{China}}}

\affil[2]{\orgname{Superbloch Capital Management Co., Ltd.}, \orgaddress{\city{Tokyo}, \postcode{101-0027}, \country{Japan}}}

\affil[3]{\orgname{Beijing Superbloch Technology Inc.}, \orgaddress{\city{Beijing}, \postcode{100082}, \country{China}}}

\affil[4]{\orgdiv{Yukawa Institute for Theoretical Physics},\orgname{Kyoto University}, \orgaddress{ \city{Kyoto}, \postcode{606-8502}, \country{Japan}}}

\affil[5]{\orgdiv{Institute of Theoretical Physics},\orgname{Chinese Academy of Sciences}, \orgaddress{ \city{Beijing}, \postcode{100190}, \country{China}}}

\affil[6]{\orgdiv{School of Physical Sciences},\orgname{University of Chinese Academy of Sciences}, \orgaddress{\street{\city{Beijing}, \postcode{100049}, \country{China}}}}

\affil[7]{\orgdiv{MinJiang Collaborative Center for Theoretical Physics},\orgname{MinJiang University}, \orgaddress{\street{\city{Fuzhou}, \postcode{350108}, \country{China}}}}


\abstract{
Fluctuation theorems show how coarse graining transforms microscopic symmetry into observable irreversibility. Here we ask whether an analogous symmetry-based diagnostic can be constructed for financial markets. At the microscopic level, each transaction pairs a buyer and a seller, whereas trading decisions are typically made from coarse-grained price histories. Using symmetric take-profit and stop-loss rules, we compare the holding-time distributions of long and short trading ensembles generated from the same price series. Across equity indices, individual stocks and cryptocurrencies, the log-ratio of the two distributions shows a robust crossover. It remains nearly constant at short durations but becomes linear in the tail, implying an exponential directional asymmetry. The tail slope defines an effective market temperature, an operational measure of fluctuation intensity on the chosen observation scale. A Bachelier first-passage benchmark captures the exponential tails but not the asymmetry, because long and short positions share the same leading decay rate. By contrast, short-time correlations between overlapping positions provide a minimal mechanism for the asymmetry by generating direction-dependent subleading relaxation spectra in a coarse-grained Markov description. Together, these results establish a fluctuation-theorem-like diagnostic of irreversibility in financial markets and, more broadly, in complex systems accessible only through coarse-grained observables.
}

\maketitle
\newpage
Fluctuation theorems (FTs) provide a quantitative framework for relating microscopic symmetry to macroscopic irreversibility in nonequilibrium systems \cite{EvansCohenMorriss1993,GallavottiCohen1995,Kurchan1998,LebowitzSpohn1999,Seifert2005,Jarzynski1997,Crooks1998,Crooks1999,EspositoVandenBroeck2010}. In their standard form, paired microscopic trajectories connected by an underlying symmetry remain admissible, while coarse graining reshapes their statistical weights into measurable asymmetries in the statistics of work, heat and entropy production. Recent work has further emphasized that these relations form a hierarchy, with irreversibility progressively transferred from resolved to hidden degrees of freedom as information is coarse-grained away \cite{KawaguchiNakayama2013,ChenQuan2023}. This perspective motivates a broader question of whether comparable symmetry-based diagnostics of coarse-grained irreversibility can be identified in complex systems beyond conventional thermodynamic settings.

\begin{figure}
    \centering
    \includegraphics[width=1.0\linewidth]{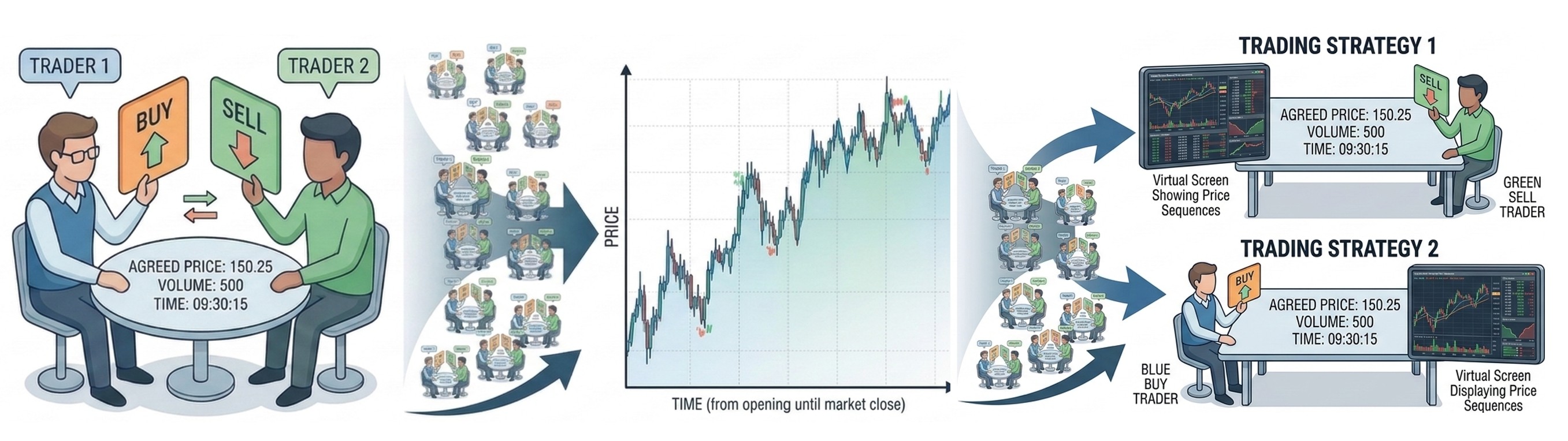}
    \caption{From microscopic buyer-seller transactions to coarse-grained market observables. At the microscopic level, the market consists of matched buyer-seller events carrying transaction-level information. After coarse graining, traders access only public market observables such as the price path and traded volume. Applying symmetric long and short trading rules to the same coarse-grained price history generates the holding-time distributions $P_{+}(d_p)$ and $P_{-}(d_p)$, whose log-ratio defines the symmetry diagnostic $\Phi(d_p)$.}
    \label{fig:concept}
\end{figure}

Financial markets provide a natural setting for this question. At the microscopic level, each transaction is a matched buyer-seller event carrying price, volume and time information \cite{Hasbrouck2007,OHara1995}. At the level of observation, however, most market participants do not access the full market state, but instead act on publicly available coarse-grained signals such as prices and traded volumes \cite{Hasbrouck2007,OHara1995}. This separation suggests an operational analogy. The microscopic pairing of buyers and sellers provides the transaction-level counterpart of an underlying symmetry, while the long and short duration statistics generated by the same trading rule on the same price history define the coarse-grained observables of interest. The logic of this construction is illustrated schematically in Fig.~\ref{fig:concept}.

The analogy should be interpreted cautiously. Markets are adaptive, strategic and nonstationary, and we do not claim an exact thermodynamic entropy-production identity. Our aim is instead to determine how strongly coarse graining breaks directional symmetry in market data. To this end, we consider the simplest first-passage observable associated with a symmetric trading rule, namely the holding time required for a long or short position to reach a prescribed take-profit or stop-loss boundary.

We therefore construct two ensembles with identical thresholds but opposite directions and compare their holding-time distributions. For a position opened at time $t_0$, we denote by $t_e$ the exit time at which either the take-profit or stop-loss boundary is first reached, and define the holding time as $d_p=t_e-t_0$. Denoting the corresponding long and short distributions by $P_{+}(d_p)$ and $P_{-}(d_p)$, we define the symmetry diagnostic
\begin{equation}\label{Eq.Ft}
\Phi(d_p)\equiv \ln \frac{P_{+}(d_p)}{P_{-}(d_p)}.
\end{equation}
If directional symmetry were preserved exactly under coarse graining, then $\Phi(d_p)$ would vanish for all $d_p$. Instead, as we show below, $\Phi(d_p)$ exhibits a robust crossover. It remains close to zero at short durations and develops an approximately linear tail at longer durations, implying an exponential directional asymmetry.

This diagnostic can now be applied directly to market data. We begin by showing how the holding time is constructed from an observed price path and then examine how the resulting asymmetry varies across time and across markets.

\section{Empirical asymmetry and an effective market temperature}

\begin{figure}
    \centering
    \includegraphics[width=1.0\linewidth]{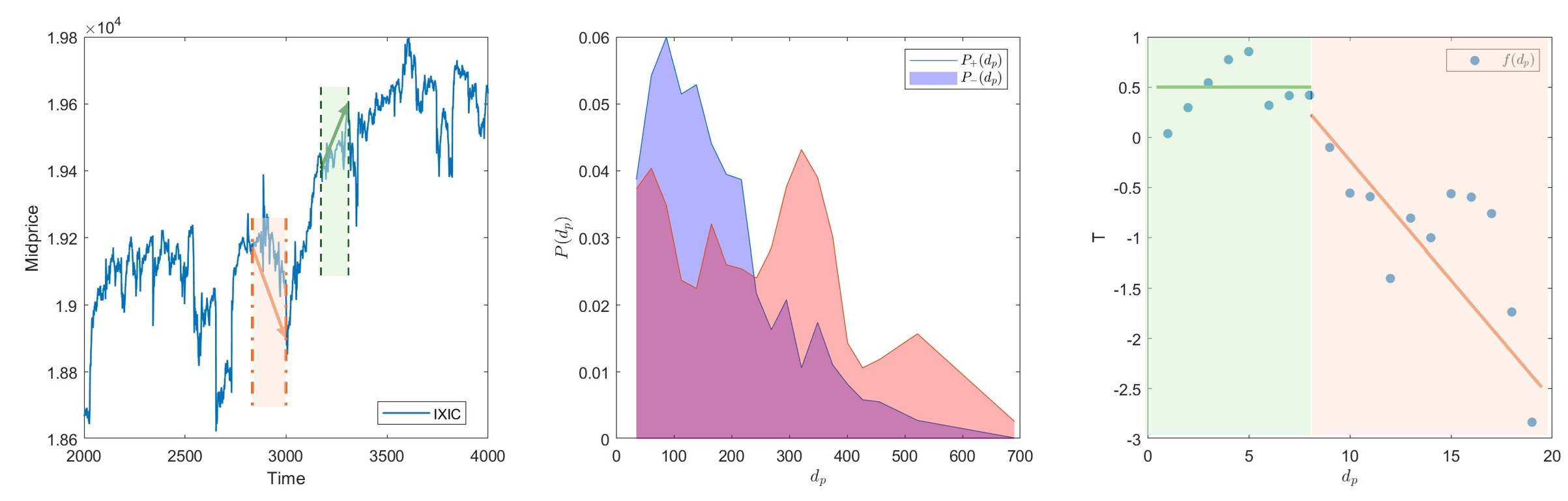}
    \caption{Construction of the holding time and the core empirical asymmetry. Panel (a) shows a representative segment of the IXIC midprice series. A sample entry time $t_{0}$ is marked together with the take-profit and stop-loss boundaries. The exit time $t_{e}$ is defined by the first threshold crossing, and the holding time is $d_p=t_{e}-t_{0}$. Panel (b) shows the empirical holding-time distributions $P_{+}(d_p)$ and $P_{-}(d_p)$ for long and short positions generated from the same price series using identical thresholds $\alpha=0.018725$ and $\beta=0.010000$. Panel (c) shows the corresponding symmetry diagnostic $\Phi(d_p)=\ln[P_{+}(d_p)/P_{-}(d_p)]$, revealing a short-duration plateau and an approximately linear tail at longer durations.}
    \label{fig:core}
\end{figure}

We first illustrate how the holding time $d_p$ is constructed from an observed price path. Figure~\ref{fig:core}(a) shows a representative segment of the midprice trajectory together with a representative entry time $t_0$, the associated take-profit and stop-loss boundaries, and the exit time $t_e$ at which one of the two thresholds is first crossed. The holding time is then defined as $d_p=t_e-t_0$. Repeating this construction over many entry times generates long and short ensembles of holding times for the same thresholds $(\alpha,\beta)$.

Using the Nasdaq Composite Index (IXIC) from April 2025 to October 2025 as an example, Fig.~\ref{fig:core}(b) shows the holding-time distributions $P_{+}(d_p)$ and $P_{-}(d_p)$ for the long and short directions. A clear directional asymmetry is already visible at the level of the distributions, and the difference becomes more pronounced as $d_p$ increases. The same asymmetry is seen more clearly in the symmetry diagnostic $\Phi(d_p)$. As shown in Fig.~\ref{fig:core}(c), $\Phi(d_p)$ exhibits a robust crossover. It remains close to zero at short durations, indicating an approximately symmetric regime, and develops an approximately linear tail at longer durations, implying an exponential directional bias between the two holding-time distributions.

This empirical structure is well captured by the piecewise form
\begin{equation}\label{Eq.FT-k}
\Phi(d_p)=
\begin{cases}
c, & d_p<d_{pc},\\[4pt]
k(d_p-d_{p0}), & d_p\ge d_{pc},
\end{cases}
\end{equation}
where $c$ is close to zero and $d_{pc}$ marks the crossover duration. The short-duration plateau corresponds to an approximately symmetric regime for rapidly closed positions, whereas the long-duration sector is clearly asymmetric. The approximately linear dependence of $\Phi(d_p)$ in this regime implies an exponential bias between the two holding-time distributions.

The slope $k$ in Eq.~\eqref{Eq.FT-k} provides a compact measure of the long-duration directional asymmetry. Motivated by the structure of fluctuation relations, we define an operational effective market temperature
\begin{equation}\label{eq:temperature}
T_{\mathrm m}:=\frac{1}{|k|},
\end{equation}
up to an arbitrary choice of units. The absolute value is introduced because the sign of $k$ depends on the relative choice of the thresholds $(\alpha,\beta)$ and can therefore change under different but equally admissible trading conventions. The quantity $|k|$, rather than the sign of $k$, measures the strength of the long-duration asymmetry. This quantity is not an equilibrium temperature. It is a coarse-grained diagnostic of fluctuation intensity on the observation scale defined by the trading rule.

\begin{figure}
    \centering
    \includegraphics[width=1.0\linewidth]{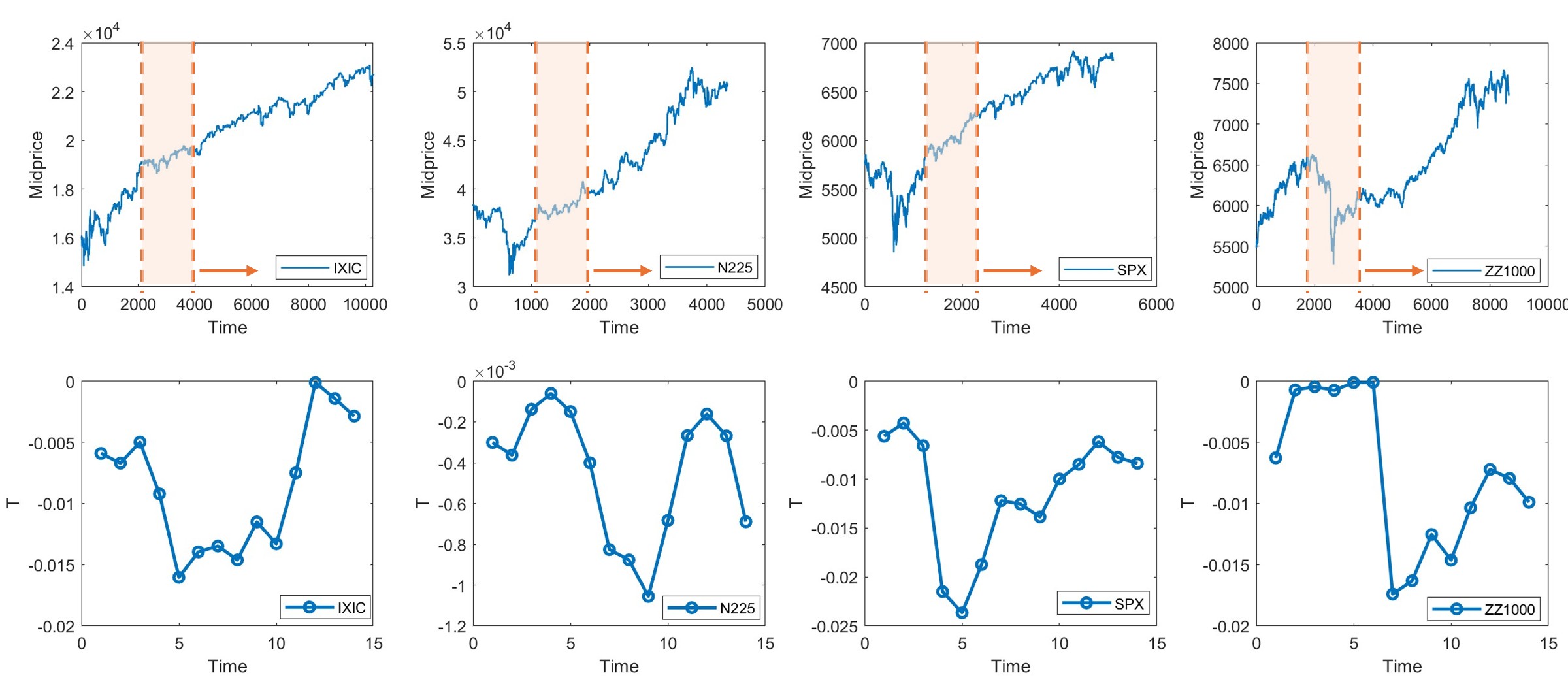}
    \caption{Temporal evolution of the effective market temperature across major global equity indices. The four columns correspond to four representative equity indices over the same analysis period. The top row shows the corresponding price trajectories, and the bottom row shows the effective market temperature $T_{\mathrm m}(t)$ extracted from the tail slope of $\Phi(d_p)$ using the same sliding-window procedure. Although the price series evolve comparatively smoothly over extended intervals, the inferred $T_{\mathrm m}(t)$ exhibits pronounced temporal variation in all four markets. This shows that the coarse-grained directional asymmetry captured by the symmetry diagnostic is not directly visible from price trajectories alone and recurs across geographically separated equity markets.}
    \label{fig:global}
\end{figure}

We next examine the temporal evolution of the effective market temperature across major global equity indices. Figure~\ref{fig:global} compares the price trajectories and the corresponding $T_{\mathrm m}(t)$ obtained from the same sliding-window procedure for four representative indices. In all four markets, the price trajectories evolve comparatively smoothly over extended intervals, whereas $T_{\mathrm m}(t)$ displays pronounced temporal structure, including sharp drops and recoveries. The directional asymmetry captured by $T_{\mathrm m}(t)$ is therefore not directly visible from price trajectories alone. Despite differences in region, volatility scale and market microstructure, substantial time dependence in $T_{\mathrm m}(t)$ appears in all four indices, indicating that the effect is a systematic feature of major global equity markets rather than an isolated property of a single index. Notably, the deepest minima occur around early July 2025 across several indices. This timing coincides with the enactment of the One Big Beautiful Bill Act\cite{OBBBAWhiteHouse2025,ReutersOBBBADebt2025} on July 4, 2025, suggesting that $T_{\mathrm m}(t)$ may be sensitive to policy-driven regime shifts that are not transparently encoded in price trajectories alone. A broader correlation analysis of $T_{\mathrm m}(t)$ across the 25 largest financial markets, presented in Supplementary B, further supports this interpretation. Over long sampling periods, most markets remain only weakly correlated, with strong persistent coupling concentrated in a few major centres such as Nasdaq and major Japanese and Chinese markets. By contrast, a windowed correlation analysis over the July--August 2025 episode reveals strong correlations across more than 18 major markets. This pattern underscores both the global reach of the event and the value of $T_{\mathrm m}(t)$ as a coarse-grained physical diagnostic of system-wide market regimes. Additional market-by-market results across a broader set of equity indices, individual stocks and cryptocurrencies are also provided in Supplementary \ref{secB}.

Taken together, these empirical results establish both the existence and the broader significance of the asymmetry. It is directly visible in the holding-time statistics generated by applying the same symmetric trading rule to long and short positions. The corresponding effective market temperature also carries temporal information that is not transparently encoded in the price series alone and recurs across multiple major markets. We now turn to the minimal stochastic mechanism capable of producing this structure.

\section{A minimal stochastic mechanism for directional asymmetry}\label{sec4}

We begin with the Bachelier model \cite{Bachelier1900}, which occupies a foundational place in mathematical finance as a minimal diffusive description of price fluctuations. Although highly idealized, it provides a clean baseline here because it isolates the interplay between stochastic diffusion and absorbing trading thresholds without additional market microstructure. We therefore use it as a representative benchmark for the mechanism analysis. Similar qualitative conclusions for other baseline price models are summarized in the Supplementary \ref{secA}.

We generate synthetic price trajectories from the arithmetic Brownian motion
\begin{equation}
dp(t)=\mu\,dt+\sigma\,dW_t,
\end{equation}
where $\mu$ and $\sigma$ denote the drift and volatility, and $W_t$ is a standard Wiener process. Using the same trading rule as in the empirical analysis, with prescribed take-profit and stop-loss thresholds $(\alpha,\beta)$, we construct long and short holding-time ensembles from the simulated trajectories. For a position opened at time $t_0$, the exit time $t_e$ is defined as the first-passage time, equivalently the first hitting time, to either threshold, and the holding time is $d_p=t_e-t_0$. The resulting distributions are again denoted by $P_{+}(d_p)$ and $P_{-}(d_p)$, so that the corresponding symmetry diagnostic $\Phi(d_p)$ can be compared directly with the empirical one.

For the independent construction, the holding-time statistics reduce to a standard first-passage problem with absorption. The corresponding distributions admit the spectral representation \cite{Risken1989,Redner2001}
\begin{equation}
P_{\pm}(d_p)=\sum_{m} A_m^{(\pm)} e^{-\lambda_m d_p},
\end{equation}
where the coefficients $A_m^{(\pm)}$ are the projections of the corresponding initial condition onto the eigenmodes of the first-passage operator. At long durations the leading mode dominates,
\begin{equation}
P_{\pm}(d_p)\sim A_{\pm}e^{-\lambda_1 d_p},
\qquad d_p\to\infty .
\end{equation}
Changing the trading direction does not alter the spectrum $\{\lambda_m\}$ of the underlying diffusive first-passage problem; it changes only the mode weights $A_m^{(\pm)}$. The two directions therefore share the same leading decay rate $\lambda_1$, and it follows that
\begin{equation}
\Phi(d_p)=\ln\frac{P_+(d_p)}{P_-(d_p)}
\longrightarrow
\ln\frac{A_+}{A_-},
\qquad d_p\to\infty .
\end{equation}
The independent Bachelier baseline therefore produces exponential tails in $P_{\pm}(d_p)$, but only a saturating asymptotic form of $\Phi(d_p)$. The mismatch with the empirically observed approximately linear tail shows that an additional mechanism beyond independent first-passage diffusion is required.

The missing ingredient is the pathwise correlation induced by sequential trading on a common price history. In the empirical analysis, positions are not drawn from independent realizations, but are opened sequentially along a single evolving trajectory, so nearby trades overlap in time and sample common local increments. To incorporate this feature without changing the microscopic Bachelier dynamics, we introduce a correlated statistic method in which holding times are measured from one long realization of the same process, exactly as in the empirical protocol. The resulting durations are therefore correlated at the ensemble level, even though the underlying price increments remain purely diffusive. For clarity, we refer to the independent first-passage construction discussed above as the uncorrelated statistic method.

\begin{figure}
    \centering
    \includegraphics[width=1\linewidth]{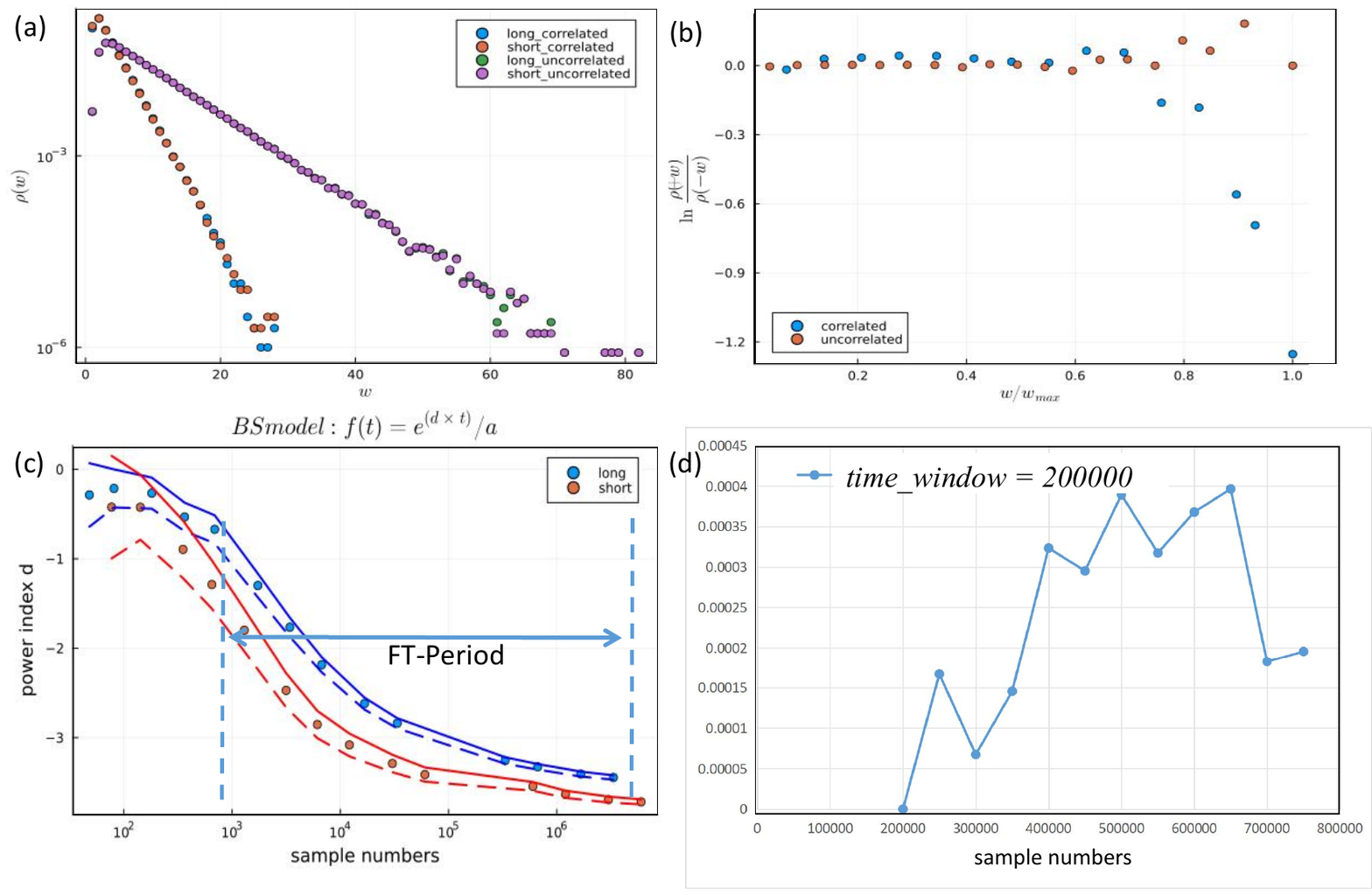}
    \caption{Bachelier benchmark with uncorrelated and correlated statistic methods. Panel (a) shows the holding-time distributions $P_{+}(d_p)$ and $P_{-}(d_p)$ for the two opposite trading directions, obtained from the same arithmetic Brownian motion using the uncorrelated and correlated ensemble constructions. In both cases, the distributions exhibit exponentially decaying tails, showing that this feature already arises from diffusion with absorbing trading thresholds. Panel (b) shows the corresponding symmetry diagnostic $\Phi(d_p)=\ln[P_{+}(d_p)/P_{-}(d_p)]$. The uncorrelated statistic method yields a long-duration behavior that bends toward saturation, whereas the correlated statistic method produces an extended approximately linear tail, in direct analogy with the empirical result. Panel (c) shows the temporal variation of the fitted slope $k$ in Eq.~\eqref{Eq.FT-k}, estimated in a moving window of 200000 samples along a single long simulated trajectory. Even in this minimal diffusive setting, the corresponding effective market temperature therefore acquires temporal structure once the coarse-grained trading ensembles retain pathwise correlations.}
    \label{fig:bachelier}
\end{figure}

These two constructions lead to qualitatively different outcomes. Figure~\ref{fig:bachelier}(a) shows that both the correlated and uncorrelated statistic methods produce exponentially decaying tails in the holding-time distributions $P_{+}(d_p)$ and $P_{-}(d_p)$, confirming that this feature already arises from diffusion with absorbing trading thresholds. The essential difference appears in the symmetry diagnostic. As shown in Fig.~\ref{fig:bachelier}(b), the uncorrelated statistic method yields a $\Phi(d_p)$ that bends toward saturation at long durations, in agreement with the independent first-passage analysis above. By contrast, the correlated statistic method produces an extended approximately linear tail in $\Phi(d_p)$, in direct analogy with the empirical result. Estimating the corresponding slope $k$ in a moving window along a single long simulated trajectory then reveals a nontrivial temporal structure, as shown in Fig.~\ref{fig:bachelier}(c). Even in this minimal diffusive setting, the corresponding effective market temperature therefore acquires time dependence once the coarse-grained trading ensembles retain pathwise correlations.

Taken together, these results show that diffusion with absorbing trading thresholds explains the common exponential baseline of the holding-time distributions, whereas the fluctuation-theorem-like asymmetry appears only when the coarse-grained trading ensembles retain pathwise correlations. The remaining task is therefore not to introduce a more elaborate price model, but to identify the minimal relaxation structure through which such correlations are converted into directional asymmetry.

The correlated Bachelier construction already shows that pathwise correlations can generate a fluctuation-theorem-like asymmetry. The next step is to identify the minimal dynamical structure responsible for this effect. To this end, we construct a Markov chain directly for the sequence of first-passage times $\{d_p^{(n)}\}$ produced by sequential trading on a common price history. The state space is the discretized holding time $d_p$ itself, so the correlation induced by the trading protocol is recast as a transition kernel acting on duration states. In this way, the mechanism can be read directly from the spectrum of the transition matrix without modifying the underlying diffusive dynamics.

\begin{figure}
    \centering
    \includegraphics[width=1\linewidth]{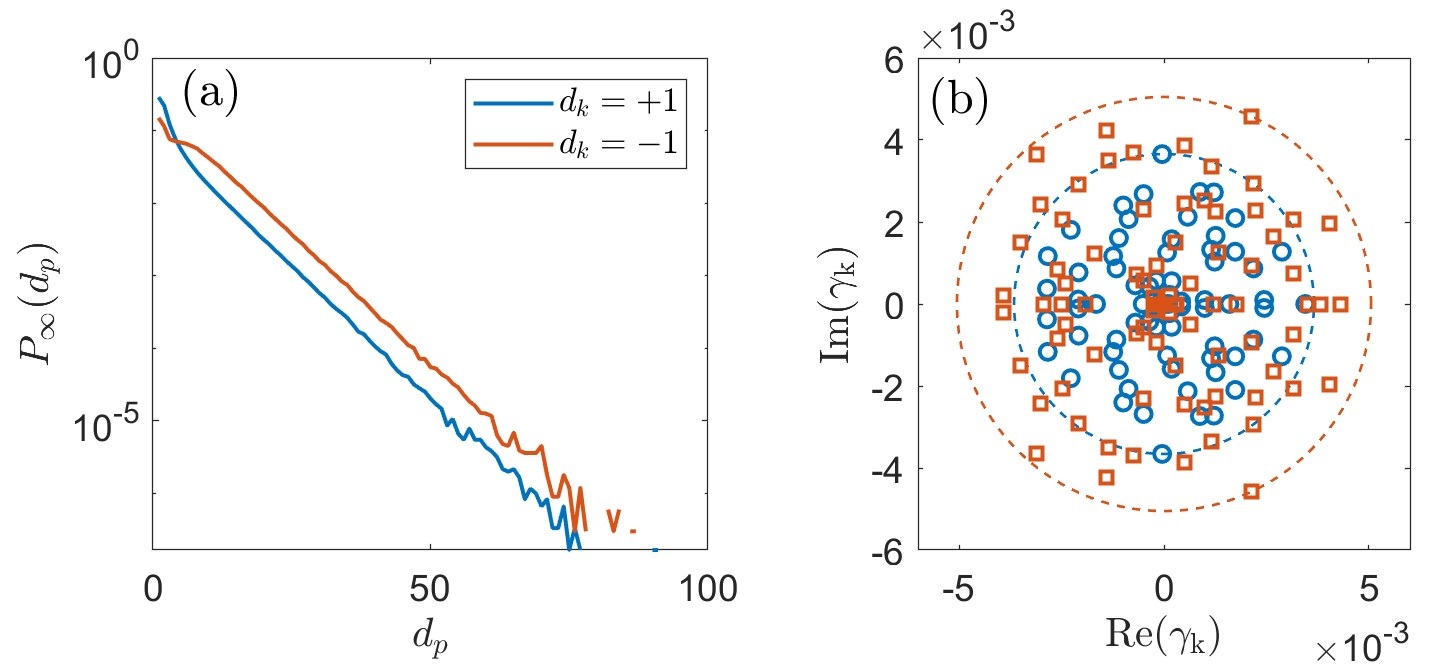}
    \caption{Markov-chain analysis of the first-passage-time sequence. The chain is constructed directly from the correlated sequence of holding times $d_p$ generated by sequential trading on a common price history, with the discretized duration $d_p$ taken as the Markov state. Panel (a) shows the stationary distributions $P_{\infty,\pm}$ for $d_k=+1$ (blue) and $d_k=-1$ (red), which define the common asymptotic baseline of the duration chain. Panel (b) shows the corresponding subleading eigenvalue spectra $\gamma_{k,\pm}$ of the transition matrices $M_{\pm}$, excluding the stationary eigenvalue $\gamma_1=1$. Although the asymptotic decay rate remains essentially direction independent, the subleading spectra differ between the two directions. Through Eq.~\eqref{Eq:Lambda-eff}, these spectral differences generate different finite-time effective decay rates and thereby explain the fluctuation-theorem-like asymmetry observed in empirical windows.}
    \label{fig:markov}
\end{figure}

Let $\mathbf{P}^{(n)}_{\pm}$ denote the probability vector over discretized holding-time states after the $n$-th trade for the two trading directions. Its evolution is described by
\begin{equation}
\mathbf{P}^{(n+1)}_{\pm}=M_{\pm}\mathbf{P}^{(n)}_{\pm},
\end{equation}
where $M_{\pm}$ is the transition matrix inferred from the correlated sequence of durations generated by the same trading protocol. Here $d_p$ is discretized using the same time resolution as in the simulated dynamics. Constructing the chain separately for $d_k=+1$ and $d_k=-1$ allows us to compare both the stationary distributions and the relaxation spectra between the two directions. Figure~\ref{fig:markov}(a) shows the stationary distributions of the duration chain, while Fig.~\ref{fig:markov}(b) shows the corresponding subleading spectra.

The directional difference appears at finite times rather than in the stationary limit. For a generic initial distribution $P_{0,\pm}(d_p)$, the duration distribution after $n$ trades admits the spectral decomposition \cite{Norris1997,LevinPeresWilmer2009}
\begin{equation}\label{Eq.Pn}
P_{n,\pm}(d_p)=P_{\infty,\pm}(d_p)+\sum_{k\ge2} c_{k,\pm}\,\gamma_{k,\pm}^{\,n}\,\nu_{k,\pm}(d_p),
\end{equation}
where $\gamma_{k,\pm}$ and $\nu_{k,\pm}(d_p)$ are the subleading eigenvalues and eigenvectors of $M_{\pm}$, and $c_{k,\pm}$ are the corresponding projections of the initial condition. As $n$ increases, all modes with $|\gamma_{k,\pm}|<1$ decay exponentially, so the distribution relaxes toward the stationary form. The fluctuation-theorem-like asymmetry is therefore a finite-time effect associated with the pre-asymptotic relaxation of the duration chain.

In practice, duration statistics are estimated over finite windows, so the relevant distribution is not fully stationary. We therefore summarize the tail behaviour of $P_{n,\pm}$ through an operational effective decay rate defined from the mean duration,
\begin{equation}
\lambda_{\mathrm{eff},\pm}(n):=\frac{1}{\mathbb{E}_{P_{n,\pm}}[d_p]}.
\end{equation}
Using Eq.~\eqref{Eq.Pn}, the mean duration can be written as
\begin{equation}
\mathbb{E}_{P_{n,\pm}}[d_p]=\frac{1}{\lambda_1}+\sum_{k\ge2} c_{k,\pm}\,\gamma_{k,\pm}^{\,n}\,\mu_{k,\pm},
\end{equation}
where $\lambda_1$ denotes the common asymptotic decay rate of the two directions and $\mu_{k,\pm}$ are the first moments of the corresponding subleading eigenmodes.

For large $n$, the contribution of the subleading modes is perturbative. Expanding $\lambda_{\mathrm{eff},\pm}(n)=1/\mathbb{E}_{P_{n,\pm}}[d_p]$ to first order around $1/\lambda_1$ gives
\begin{equation}\label{Eq:Lambda-eff}
\lambda_{\mathrm{eff},\pm}(n)\approx \lambda_1-\lambda_1^{\,2}\sum_{k\ge2} c_{k,\pm}\,\gamma_{k,\pm}^{\,n}\,\mu_{k,\pm},
\end{equation}
up to higher-order terms in the subleading modes. This expression shows explicitly that the finite-time effective decay rate is controlled by the same pre-asymptotic spectral contributions that govern the relaxation of the duration chain.

Reversing the trading direction changes the short-time correlation structure of the duration sequence and therefore modifies the effective transition kernel $M_{\pm}$. While the common asymptotic decay rate $\lambda_1$ remains essentially unchanged, the subleading spectra $\{\gamma_{k,\pm}\}_{k\ge2}$ are sensitive to these pathwise correlations and can differ between long and short directions, as shown in Fig.~\ref{fig:markov}(b). Through Eq.~\eqref{Eq:Lambda-eff}, such spectral differences generate different finite-time effective decay rates for the two directions and thereby produce a systematic bias between $P_{+}(d_p)$ and $P_{-}(d_p)$ over the range of durations accessible in empirical windows. The observed fluctuation-theorem-like scaling is therefore not a stationary property of the duration chain, but a pre-asymptotic consequence of an asymmetric relaxation spectrum induced by coarse graining.

Reversing the trading direction changes the short-time correlation structure of the duration sequence and therefore modifies the effective transition kernel $M_{\pm}$. While the common asymptotic decay rate $\lambda_1$ remains essentially unchanged, the subleading spectra $\{\gamma_{k,\pm}\}_{k\ge2}$ are sensitive to these pathwise correlations and can differ between long and short directions, as shown in Fig.~\ref{fig:markov}(b). Through Eq.~\eqref{Eq:Lambda-eff}, such spectral differences generate different finite-time effective decay rates for the two directions and thereby produce a systematic bias between $P_{+}(d_p)$ and $P_{-}(d_p)$ over the range of durations accessible in empirical windows. The observed fluctuation-theorem-like scaling is therefore not a stationary property of the duration chain, but a pre-asymptotic consequence of an asymmetric relaxation spectrum induced by coarse graining.

This mechanism also clarifies the meaning of the empirical asymmetry. It does not arise from an explicit breaking of the microscopic pairing of buyers and sellers in individual transactions, but from the way coarse graining reorganizes correlated trajectory segments into long and short trading ensembles. Financial markets therefore provide a simple setting in which microscopic transactional pairing coexists with macroscopic statistical asymmetry, and in which the latter can be traced to a finite-time asymmetric relaxation spectrum.

\section{Discussion}

The results reported here reveal a fluctuation-theorem-like asymmetry in the coarse-grained holding-time statistics of financial markets. Starting from the microscopic pairing of buyers and sellers in individual transactions, we construct long and short trading ensembles from the same price history and compare their duration distributions under the same symmetric trading rule. The resulting symmetry diagnostic $\Phi(d_p)$ remains nearly constant at short durations but develops an approximately linear tail at longer durations. This structure is observed across equity indices, individual stocks and cryptocurrencies, and is naturally summarized by the slope parameter $k$ in Eq.~\eqref{Eq.FT-k}.

The mechanism identified here is not a direct microscopic breaking of the transactional pairing of buyers and sellers, but the short-time correlation structure generated by dense entry times and overlapping positions. Such persistence is consistent with the long-memory properties of order flow and market impact documented in high-frequency financial data \cite{BouchaudGefenPottersWyart2004,BouchaudFarmerLillo2008}. Once holding times are treated as a correlated stochastic sequence, their dynamics can be represented by an effective transition operator acting on the duration distribution \cite{Norris1997,LevinPeresWilmer2009}. The asymptotic decay rate remains essentially the same in the two directions, but the subleading relaxation spectrum is sensitive to direction-dependent correlations. These spectral differences control finite-window statistics and provide a concrete route from pathwise correlations to the observed long-duration asymmetry. In this sense, the fluctuation-theorem-like scaling is a coarse-graining effect, closely analogous to hidden entropy production in stochastic thermodynamics \cite{SeifertRPP2012,CelaniEtAl2012,KawaguchiNakayama2013,ChenQuan2023}.

The slope parameter $k$ in Eq.~\eqref{Eq.FT-k} provides a compact measure of the long-duration directional bias. Writing $T_{\mathrm m}=1/|k|$ as an effective market temperature emphasizes its role as a coarse-grained measure of fluctuation intensity and irreversibility strength, rather than as a thermodynamic temperature in equilibrium. This interpretation is motivated by the general structure of fluctuation relations, in which exponential weights encode dissipation-related functionals \cite{EvansCohenMorriss1993,GallavottiCohen1995,Seifert2005,EspositoVandenBroeck2010}. We stress that $T_{\mathrm m}$ is an operational diagnostic defined by a particular coarse-graining procedure, observation scale and trading rule. It should not be taken to imply equilibrium, detailed balance, or a Boltzmann-Gibbs description of the market. At the same time, the definition of $k$ is model independent at the level of the empirical diagnostic. It depends only on the long and short duration statistics generated by a fixed symmetric trading rule, and can therefore be estimated consistently across assets and time periods.

Several limitations and possible extensions deserve mention. The precise values of $(\alpha,\beta)$ affect the range of durations over which the crossover and tail can be resolved with sufficient statistics, and therefore influence the stability of the fitted slope. More generally, the present construction focuses on holding time as a coarse-grained observable, whereas other microstructural variables, including signed order flow, volume imbalance and realized impact, may carry complementary signatures of irreversibility \cite{Hasbrouck2007,OHara1995,Kyle1985,BouchaudFarmerLillo2008}. In addition, although the Markov-operator framework provides a minimal route for incorporating correlations, the true duration dynamics may exhibit higher-order memory. Determining the effective state space required to recover directional symmetry remains an open question. On the empirical side, it would also be valuable to study more systematically how the inferred asymmetry responds to changes in liquidity, trading conventions and major macro-financial events.

Several limitations and possible extensions deserve mention. The precise values of $(\alpha,\beta)$ affect the range of durations over which the crossover and tail can be resolved with sufficient statistics, and therefore influence the stability of the fitted slope. More generally, the present construction focuses on holding time as a coarse-grained observable, whereas other microstructural variables, including signed order flow, volume imbalance and realized impact, may carry complementary signatures of irreversibility \cite{Hasbrouck2007,OHara1995,Kyle1985,BouchaudFarmerLillo2008}. In addition, although the Markov-operator framework provides a minimal route for incorporating correlations, the true duration dynamics may exhibit higher-order memory. Determining the effective state space required to recover directional symmetry remains an open question. On the empirical side, it would also be valuable to study more systematically how the inferred asymmetry responds to changes in liquidity, trading conventions and major macro-financial events.

\begin{appendices}

\section{Separation-of-Variables Derivation}\label{secA}

\subsection{Biased diffusion with absorbing boundaries}
We consider the biased diffusion (Fokker--Planck) equation
\begin{equation}
    \frac{\partial c(p,t)}{\partial t}+\mu\,\frac{\partial c(p,t)}{\partial p}=D\,\frac{\partial^2 c(p,t)}{\partial p^2},\qquad p\in(0,L),\ t>0,
\end{equation}
with absorbing boundary conditions
\begin{equation}
    c(0,t)=c(L,t)=0,
\end{equation}
and point-source initial condition
\begin{equation}
    c(p,0)=\delta(p-p_0),\qquad p_0\in(0,L).
\end{equation}

\subsection{Separation of variables}
We seek separated solutions of the form $c(p,t)=f(t)g(p)$. Substituting into the PDE gives
\begin{equation}
    f'(t)g(p)+\mu f(t)g'(p)=Df(t)g''(p).
\end{equation}
Dividing by $f(t)g(p)$ yields
\begin{equation}
    \frac{f'(t)}{f(t)}=\frac{Dg''(p)-\mu g'(p)}{g(p)}=-\lambda,
\end{equation}
where $\lambda$ is the separation constant. Therefore,
\begin{equation}
    f(t)=A\,e^{-\lambda t},
\end{equation}
and $g$ satisfies the ODE
\begin{equation}\label{Eq:g_ode}
    Dg''(p)-\mu g'(p)+\lambda g(p)=0,
\end{equation}
with $g(0)=g(L)=0$.

\subsection{Spatial eigenfunctions and eigenvalues}
Assuming $g(p)=e^{\beta p}$ leads to the characteristic equation
\begin{equation}
    D\beta^2-\mu\beta+\lambda=0.
\end{equation}
A nontrivial solution satisfying Dirichlet boundaries requires an oscillatory mode. It is convenient to factor out the exponential bias and write
\begin{equation}
    g(p)=e^{\frac{\mu}{2D}p}\,\phi(p).
\end{equation}
Substituting into Eq.~\eqref{Eq:g_ode} yields
\begin{equation}
    \phi''(p)+\kappa^2\phi(p)=0,\qquad \kappa^2:=\frac{\lambda}{D}-\frac{\mu^2}{4D^2}.
\end{equation}
With $g(0)=g(L)=0$ (hence $\phi(0)=\phi(L)=0$), we obtain
\begin{equation}
    \phi_n(p)=\sin\!\left(\frac{n\pi p}{L}\right),\qquad \kappa_n=\frac{n\pi}{L},\qquad n=1,2,\dots
\end{equation}
and therefore the eigenvalues
\begin{equation}\label{Eq:lambda_n_appendix}
    \lambda_n=\frac{\mu^2}{4D}+D\left(\frac{n\pi}{L}\right)^2.
\end{equation}
The corresponding eigenfunctions are
\begin{equation}
    g_n(p)=e^{\frac{\mu}{2D}p}\sin\!\left(\frac{n\pi p}{L}\right).
\end{equation}

\subsection{Superposition and the point-source initial condition}
We write the solution as a superposition
\begin{equation}
    c(p,t)=\sum_{n=1}^{\infty} E_n\,e^{\frac{\mu}{2D}p}\sin\!\left(\frac{n\pi p}{L}\right)e^{-\lambda_n t}.
\end{equation}
At $t=0$ this must equal $\delta(p-p_0)$. Multiplying both sides by $e^{-\frac{\mu}{2D}p}\sin\!\left(\frac{m\pi p}{L}\right)$ and integrating over $(0,L)$, orthogonality gives
\begin{equation}
    E_n=\frac{2}{L}\,e^{-\frac{\mu}{2D}p_0}\sin\!\left(\frac{n\pi p_0}{L}\right).
\end{equation}
Substituting $E_n$ into the superposition yields:
\begin{equation}
 c(p, t)=\frac{2}{L} \sum_{n=1}^{\infty} e^{\frac{\mu\left(p-p_{0}\right)}{2 D}} \sin \frac{n \pi p_{0}}{L} \sin \frac{n \pi p}{L} e^{-\lambda_n t}.
\end{equation}

\subsection{First-passage-time density as boundary flux}
Define the probability current $J(p,t)=\mu c(p,t)-D\,\partial_p c(p,t)$. The first-passage-time density equals the total outward flux through the two absorbing boundaries,
\begin{equation}
    P(t)=J(0,t)-J(L,t).
\end{equation}
Because $c(0,t)=c(L,t)=0$, this reduces to
\begin{equation}
    P(t)=-D\,\partial_p c(p,t)\big|_{p=0}+D\,\partial_p c(p,t)\big|_{p=L}.
\end{equation}
Evaluating $\partial_p c$ and using $\cos(0)=1$ and $\cos(n\pi)=(-1)^n$ gives
\begin{equation}
    \begin{aligned}
    P(t)=&\frac{2\pi D}{L^2} \exp\!\left(-\frac{\mu p_0}{2D}\right) \sum_{n=1}^{\infty} n \sin \left( \frac{n \pi p_0}{L} \right) \\
    &\times\left[ (-1)^{n+1} \exp\!\left(\frac{\mu L }{2D}\right) +1 \right] e^{-\lambda_n t},
    \end{aligned}
\end{equation}

\section{Procedure of Fluctuation-Theorem Analysis of GOOGL Intraday Data}\label{secB}

We analyze intraday price fluctuations of \texttt{GOOGL} by testing an empirical fluctuation-theorem (FT) type symmetry on first-passage statistics derived from rolling windows of market data downloaded via \emph{Yahoo Finance}. For each window of length $T_w=\SI{30}{day}$ sampled at $\Delta t=\SI{5}{min}$ (timestamps expressed in the $+08{:}00$ timezone), we construct a family of symmetric forward--reverse trading experiments parameterized by stop-loss $s$ and take-profit $\tau$. At every admissible entry time $t_{\mathrm{entry}}$, the observed price path is shifted by a constant offset
\begin{equation}
    \Delta p = p_{\mathrm{ref}} - p(t_{\mathrm{entry}}), 
    \qquad 
    p^{(\mathrm{shift})}(t) = p(t) + \Delta p,
\end{equation}
so that all trajectories share the same reference level $p_{\mathrm{ref}}$ and the same absolute barrier magnitudes can be applied consistently across entry points.

We then define a \emph{forward} process (long) and a \emph{reverse} process (short) using identical barrier magnitudes $(s,\tau)$, and record the signed first-passage index displacement
\begin{equation}
    w = i_{\mathrm{hit}} - i_{\mathrm{entry}},
\end{equation}
with $w>0$ for forward realizations and $w<0$ for reverse realizations under the corresponding sign convention. Pooling across all entry points within the window yields empirical distributions $P_F(w)$ and $P_R(w)$ (optionally after symmetric histogram binning), from which we compute the FT diagnostic
\begin{equation}
    L(w) \equiv \ln\frac{P_F(w)}{P_R(-w)}.
\end{equation}
To isolate an asymptotic FT-like regime while accounting for short-$|w|$ deviations, we fit a piecewise model consisting of (i) a plateau for $w \le w_{t_0}$ and (ii) a linear tail for $w \ge w_{t_0}$,
\begin{equation}
    L(w)\approx k\,w+b, \qquad (w\ge w_{t_0}),
\end{equation}
where the breakpoint $t_0$ is selected from a prescribed range by maximizing a goodness-of-fit objective (e.g., $R^2$) on the tail. Finally, we optimize $(s,\tau)$ over a specified grid by selecting the configuration that maximizes the piecewise-fit objective, and report the best-fit slope $k$ (interpretable as a time-asymmetry/irreversibility strength for that window) together with the corresponding $R^2$. Results are visualized by plotting $L(w)$ with the fitted plateau/linear segments alongside the underlying close-price series to contextualize the inferred symmetry with the contemporaneous market regime.

An especially notable episode appears around early July 2025, when all four major equity indices in Fig.~\ref{fig:global} show pronounced minima in the effective market temperature $T_{\mathrm m}(t)$. This timing coincides with the enactment of the One Big Beautiful Bill Act on July 4, 2025.\cite{OBBBAWhiteHouse2025} Contemporary market commentary linked the bill to concerns about fiscal deficits, the debt limit, future Treasury issuance and the repricing of long-duration government bonds.\cite{ReutersOBBBADebt2025} We emphasize that the present analysis does not establish a causal attribution. Rather, the temporal alignment suggests that the coarse-grained symmetry diagnostic may be sensitive to macro-financial regime shifts whose effects are not directly visible in the price trajectories alone.




\end{appendices}



\begin{thebibliography}{99}

% --- Introduction: fluctuation theorems ---
\bibitem{EvansCohenMorriss1993}
D. J. Evans, E. G. D. Cohen, and G. P. Morriss,
Probability of second law violations in shearing steady states,
Phys. Rev. Lett. \textbf{71}, 2401 (1993).

\bibitem{GallavottiCohen1995}
G. Gallavotti and E. G. D. Cohen,
Dynamical ensembles in nonequilibrium statistical mechanics,
Phys. Rev. Lett. \textbf{74}, 2694 (1995).

\bibitem{Seifert2005}
U. Seifert,
Entropy production along a stochastic trajectory and an integral fluctuation theorem,
Phys. Rev. Lett. \textbf{95}, 040602 (2005).

\bibitem{Jarzynski1997}
C. Jarzynski,
Nonequilibrium equality for free energy differences,
Phys. Rev. Lett. \textbf{78}, 2690 (1997).

\bibitem{Crooks1998}
G. E. Crooks,
Nonequilibrium measurements of free energy differences for microscopically reversible Markovian systems,
J. Stat. Phys. \textbf{90}, 1481 (1998).

\bibitem{Crooks1999}
G. E. Crooks,
Entropy production fluctuation theorem and the nonequilibrium work relation for free energy differences,
Phys. Rev. E \textbf{60}, 2721 (1999).

\bibitem{Kurchan1998}
J. Kurchan,
Fluctuation theorem for stochastic dynamics,
J. Phys. A \textbf{31}, 3719 (1998).

\bibitem{LebowitzSpohn1999}
J. L. Lebowitz and H. Spohn,
A Gallavotti--Cohen-type symmetry in the large deviation functional for stochastic dynamics,
J. Stat. Phys. \textbf{95}, 333 (1999).

\bibitem{EspositoVandenBroeck2010}
M. Esposito and C. Van den Broeck,
Three detailed fluctuation theorems,
Phys. Rev. Lett. \textbf{104}, 090601 (2010).

% --- Introduction: hierarchy / coarse graining ---
\bibitem{ChenQuan2023}
J.-F. Chen and H.-T. Quan,
Hierarchical structure of fluctuation theorems for a driven system in contact with multiple heat reservoirs,
Phys. Rev. E \textbf{107}, 024135 (2023).

\bibitem{KawaguchiNakayama2013}
K. Kawaguchi and Y. Nakayama,
Fluctuation theorem for hidden entropy production,
Phys. Rev. E \textbf{88}, 022147 (2013).

% --- Introduction: market microstructure ---
\bibitem{Hasbrouck2007}
J. Hasbrouck,
\textit{Empirical Market Microstructure: The Institutions, Economics, and Econometrics of Securities Trading}
(Oxford University Press, Oxford, 2007).

\bibitem{OHara1995}
M. O'Hara,
\textit{Market Microstructure Theory}
(Blackwell, Oxford, 1995).

\bibitem{Kyle1985}
A. S. Kyle,
Continuous auctions and insider trading,
Econometrica \textbf{53}, 1315 (1985).

\bibitem{BouchaudFarmerLillo2008}
J.-P. Bouchaud, J. D. Farmer, and F. Lillo,
How markets slowly digest changes in supply and demand,
in \textit{Handbook of Financial Markets: Dynamics and Evolution},
edited by T. Hens and K. R. Schenk-Hopp\'e
(Elsevier, 2009);
also available as arXiv:0809.0822.

% --- Section 4, first subsection: Bachelier and first-passage ---
\bibitem{Bachelier1900}
L. Bachelier,
Th\'eorie de la sp\'eculation,
Ann. Sci. \'Ecole Norm. Sup. \textbf{17}, 21 (1900).

\bibitem{Risken1989}
H. Risken,
\textit{The Fokker--Planck Equation: Methods of Solution and Applications}
(Springer, Berlin, 1989).

\bibitem{Redner2001}
S. Redner,
\textit{A Guide to First-Passage Processes}
(Cambridge University Press, Cambridge, 2001).

\bibitem{BrayMajumdarSchehr2013}
A. J. Bray, S. N. Majumdar, and G. Schehr,
Persistence and first-passage properties in nonequilibrium systems,
Adv. Phys. \textbf{62}, 225 (2013).

% --- Section 4, second subsection: order-flow memory + coarse-graining ST ---
\bibitem{BouchaudGefenPottersWyart2004}
J.-P. Bouchaud, Y. Gefen, M. Potters, and M. Wyart,
Fluctuations and response in financial markets: the subtle nature of ``random'' price changes,
Quant. Finance \textbf{4}, 176 (2004).

\bibitem{SeifertRPP2012}
U. Seifert,
Stochastic thermodynamics, fluctuation theorems and molecular machines,
Rep. Prog. Phys. \textbf{75}, 126001 (2012).

\bibitem{CelaniEtAl2012}
A. Celani, S. Bo, R. Eichhorn, and E. Aurell,
Anomalous thermodynamics at the microscale,
Phys. Rev. Lett. \textbf{109}, 260603 (2012).

% --- Section 4, second subsection: Markov chains and spectral relaxation ---
\bibitem{Norris1997}
J. R. Norris,
\textit{Markov Chains}
(Cambridge University Press, Cambridge, 1997).

\bibitem{Gardiner2009}
C. W. Gardiner,
\textit{Stochastic Methods: A Handbook for the Natural and Social Sciences},
4th ed.
(Springer, Berlin, 2009).

\bibitem{LevinPeresWilmer2009}
D. A. Levin, Y. Peres, and E. L. Wilmer,
\textit{Markov Chains and Mixing Times}
(American Mathematical Society, Providence, 2009).

\bibitem{OBBBAWhiteHouse2025}
The White House,
\textit{President Trump's One Big Beautiful Bill Is Now the Law}
(The White House, Washington, DC, 2025).

\bibitem{ReutersOBBBADebt2025}
Reuters,
\textit{Trump tax bill averts one debt crisis but makes future financial woes worse}
(Reuters, July 3, 2025).



\end{thebibliography}
\end{document}